\documentclass{ws-fnl}
\usepackage{cite}
\begin{document}

\markboth{P. Bordone, F. Buscemi, C. Benedetti}{Effect of Markov and non-Markov Classical Noise on 
Entanglement Dynamics}


\title{EFFECT OF MARKOV AND NON-MARKOV CLASSICAL NOISE\\ 
ON ENTANGLEMENT DYNAMICS}

\author{\footnotesize PAOLO BORDONE}

\address{Dipartimento di Fisica, Universit\`a di Modena e Reggio Emilia, \\
and Centro S3, CNR-Istituto di Nanoscienze, \\ Via Campi 213/A,
Modena, I-41125, Italy \\
bordone@unimore.it}

\author{FABRIZIO BUSCEMI}

\address{ARCES e Dipartimento di Elettronica, Informatica e Sistemi, Universit\`a di Bologna, \\
Viale Toffano 2/2, Bologna, I-40125, Italy \\
fabrizio.buscemi@unimore.it}

\author{CLAUDIA BENEDETTI}

\address{Dipartimento di Fisica, Universit\`a degli Studi di Milano\\
Via Celoria 16, Milano, I-20133, Italy \\
claudia.benedetti@unimi.it}

\maketitle

\begin{history}
\received{(received date)}
\revised{(revised date)}
\end{history}

\begin{abstract}
We analyze the effect of a classical noise into the entanglement dynamics
between two particles, initially entangled, subject to continuous time quantum walks in a 
one-dimensional lattice. The noise is modeled by randomizing the transition amplitudes
from one site to another. Both Markovian and non-Markovian environments are 
considered. For the Markov regime an exponential decay of the initial quantum correlation is
found, while the loss of coherence of the quantum state increases monotonically with time
up to a saturation value depending upon the degrees of freedom of the system.
For the non-Markov regime the presence or absence of entanglement revival and entanglement 
sudden death phenomena is found or deduced depending on the peculiar characteristics of the noise.
Our results indicate that the entanglement dynamics in the non-Markovian regime 
is affected by the persistence of the memory effects of the environment and by its 
intrinsic features.

\end{abstract}

\section{Introduction}

Quantum entanglement represents undoubtedly one of the most peculiar aspects of quantum mechanics 
as it can be viewed as the furthest departure of the quantum world from the classical one~\cite{1}. 
Furthermore, in the last years, entanglement has been recognized as the fundamental resource for 
quantum information processing and communication, such as quantum cryptography, teleportation and 
exponential speed-up of specific computational tasks~\cite{2,3}.
The single-system decoherence, due to the environment, can make the entanglement disappear on the 
short-range time scale, making the quantum parallelism, essential for quantum computation, 
ineffective~\cite{4}. On the other hand, environment can even resume  entanglement  or  preserve it.
Thus, it is still very important to analyze the effect of the various kinds of environmental noise 
on the entanglement dynamics in realistic quantum systems which can exhibit peculiar phenomena, 
such as entanglement sudden death (ESD), and entanglement revival (ER)~\cite{5}.

ESD, namely  the disappearance of the quantum correlations between  two quantum
systems at finite time in spite 
of an exponential decay of the system coherence, appears to be peculiar of the Markov environments,
namely environments with short, or rather instantaneous, self-correlations~\cite{6}. Instead, ER,
that is the rise up of  quantum correlations in a bipartite system after a finite time period when 
they completely disappear,
occurs in presence of non-Markovian noise~\cite{7}, which represents the dominant source of decoherence 
in solid-state systems. The presence of environments with memory~\cite{7} is crucial for 
the non-monotonic time dependence of the entanglement which cannot be ascribed to 
interaction between the two sub-systems. 
Indeed, ER is found  both for couples of quantum systems interacting directly or indirectly in a common  
quantum reservoir and for  noninteracting quantum systems in independent 
non-Markovian quantum environments.

The effect of the quantum noise on the entanglement dynamics between two quantum systems
has been interpreted in terms of the transfer of the correlations back and forth from the systems 
themselves
to the environment. This is due to the back-action of the systems
on the environment. However, recent works showed that the non-monotonic time dependence
of the amount of quantum correlations may occur in two-qubit systems under the local action of
a system-unaffected environments, such as  classical random external potentials~\cite{8,9}.
Due to the classical nature of the noise, in this case no back-action-induced correlations 
can be transferred from the system to the environment.  
Thus, the occurrence of ER in a non-Markovian classical environment  
is in contrast to the well-established interpretation of revivals in terms of system-environment 
quantum back action and rises the fundamental question of how one could explain the effect of 
classical noise on entanglement dynamics in bipartite systems.

In this work, we intend to analyze the role played by a classical noisy environment
into the entanglement dynamics between two quantum particles in a simple model,
which not only allows us to relate the intrinsic features of the two-particle dynamics to the revival of 
the correlations, but can also  be of great relevance  in various physical phenomena. The diffusion of 
two quantum particles along a one-dimensional  lattice in the presence of static and dynamic disorder,
mimicked by a noisy potential, represents a good candidate for our theoretical
investigation.
In fact,  the spreading of a quantum-mechanical
wave packet in a tight-binding model with a stochastic potential is able to describe
electronic transport in solid-state systems, particle
diffusion in molecular crystals, diffusion of excitations, photon propagation
in coupled waveguide lattices. Furthermore, it  gives the possibility to connect
the transition from quantum to classical behavior of the two particles
to the features of the environmental noise (correlation time), 
as suggested by investigations on  the diffusion of single-particle wavepackets~\cite{10}.
This connection could be viewed  as  a valid guideline to study in details,
and, possibly, to clarify how Markov and non-Markov classical noise
affect the occurrence of ESD and ER.

In our approach, the dynamics of the two-particles subject to
a random potential will be faced numerically. 
The effect of noise will be simulated both by assuming
random and time-independent amplitudes for the  transitions from one site to another and
by using a sort of dynamic disorder, expressed by a potential term describing a random telegraph noise.

The paper is organized as follows. The physical model adopted is described in Sec. 2. In Sec. 3
are presented the results for the various cases considered: static and dynamic noise, Markovian 
and non-Markovian environments. Conclusions, comments and future perspectives are given in 
Sec. 4.

\section{Physical Model}

A model of two distinguishable particles continuous-time quantum walks (CTQW) in 1D graphs is 
solved by means of numerical techniques and the  time evolution  of the wavefunction 
describing the system is then used to estimate the  degree of quantum correlations.

We study CTQW on a 1D ring lattices of $N$ sites
(with $N$ even) with periodic boundary conditions.
In agreement with previous single-particle investigations~\cite{11},
the topology of the graphs here considered is simple, in the sense that each node is connected
to its two first neighbors, still it is good enough to describe experimental 
implementations of  CTQW, such as  two-photon transport in an array of waveguide lattices 
where non-classical correlations appear.
Therefore our model represents a valid tool to analyze the amount of 
quantum correlations appearing in two-particle CTQW.

The  two-particle Hamiltonian describing the dynamical evolution of  the system  
is given by
\begin{equation} \label{des}
 \mathcal{H} =\mathcal{H}^0_A+\mathcal{H}^0_B,
\end{equation}
where $\mathcal{H}^0_{A(B)}$ is the single-particle Hamiltonian  acting on 
the particle $A(B)$:
\begin{equation} \label{single}
\mathcal{H}^0_{A(B)} = \sum_{j=1}^{N}\left\{\beta_j^{A(B)}\ |j\rangle_{A(B)}\langle j|
- \ c_j^{A(B)}(t) \left(|{j+1}\rangle_{A(B)}\langle j|
+|j\rangle_{A(B)}\langle{j+1}|\right)\right\}.
\end{equation}
The kets $|j\rangle_{A(B)}$ indicate the quantum states describing the particle $A(B)$ 
localized in the $j$ node and form a complete, orthonormalized basis set, which span the 
whole accessible Hilbert space. $\beta_j^{A(B)}$ is the on-site energy, and 
$c_j^{A(B)}(t)$ is the tunneling amplitude between nearest neighbors\cite{12}. 
Due to the periodic boundary conditions, here we assume that the site $N+1$ coincides 
with site 1.

To evaluate the dynamics of the entanglement between the two particles and of 
the decoherence induced by the noisy environment on the two-particle state, we need 
to calculate the density matrix of the system as a function of time. 
To this aim we need to estimate
the evolution operator of the system which is a function of the eigenvectors and eigenvalues
of the Hamiltonian which, in turns, depends on the parameters  $\beta_j^{A(B)}$ and 
$c_j^{A(B)}(t)$. Therefore, once a suitable choice of such parameters is done, we 
diagonalize numerically the Hamiltonian for specific times and get the corresponding 
density matrix values.

In order to describe the effect of the environment on the two-particle quantum state,
one can introduce noise either through the coupling constant $\beta_j^{A(B)}$ or 
through $c_j^{A(B)}(t)$. In agreement with previous investigations\cite{12,13}, we adopt the latter.
Once noise has been included in the model, the state of the system is represented
by a density matrix $\langle\rho(t)\rangle$ which is the result of an average over a 
number of density matrices each one obtained with a specific choice of the $c_j^{A(B)}(t)$'s.
In particular, to estimate $\langle\rho(t)\rangle$, first we evaluate 
a single numerical \emph{run} by generating a sequence
of  transitions with random amplitudes  and then  by solving exactly the two-particle
Schr\"{o}dinger equation for that given sequence. The final numerical simulation
is found by producing a large number of \emph{runs}, each with a different sequence,
and then calculating the average density matrix over all the \emph{runs}.

Decoherence, which is an evaluator of the degree of entanglement between the 
two-particle system and the noisy environment, is here estimated by means of the 
von Neumann entropy\cite{14} of $\langle\rho(t)\rangle$:
\begin{equation} \label{vN}
\epsilon_{vN}(t)=-\textrm{Tr}\left\{\langle\rho(t)\rangle\ln\langle\rho(t)\rangle\right\}\ .
\end{equation}
Note that such calculation implies the numerical diagonalization of the 
$N\times N$ matrix $\langle\rho\rangle$, that becomes more and more
demanding with increasing the number of sites of the graph.

The particle-particle entanglement is quantified in terms of the negativity\cite{15}:
\begin{equation} \label{Neg}
\mathcal{N}=\sum_i\left|\lambda_i\left(\langle\rho^{T_{A(B)}}\rangle\right)\right|-1\ ,
\end{equation}
with $\lambda_i\left(\langle\rho^{T_{A(B)}}\rangle\right)$ the eigenvalues of
$\langle\rho^{T_{A(B)}}\rangle$ which is the partial transpose related to
the subsystems $A(B)$ of the total density matrix $\langle\rho\rangle$. Negativity
has been shown to be a valid measure of the degree of quantum correlations only
between two quantum sub-systems each having two degrees of freedom, namely
qubits. In fact no measure discriminating separable from entangled states for a 
mixed state with more than two degrees of freedom is known. 
As a matter of fact, non-zero negativity, for the general
case of $N$ degrees of freedom, is just a sufficient condition for entanglement\cite{16}.

\section{Numerical results}
In this section we present the numerical results obtained for the case of 
$N=4,6,8$ sites for two different ways of modeling classical noise, namely
through static and dynamic
disorder. In particular the former is simulated by assuming random and 
time-independent values for each $c_j^{A(B)}$, while the latter is modeled by
selecting the $c_j^{A(B)}(t)$ according to a random telegraph signal.

\subsection{Static noise}
In the static disorder case, the parameters $c_j^{A(B)}$'s are assumed to be
random variables following the flat probability distribution $P(c)=\frac{1}{\Delta}$ for 
$|c-c_0|\le\frac{\Delta}{2}$ and 0 otherwise\cite{13}. $c_0$ is the mean value of the 
distribution and $\Delta$ is a measure of the disorder of the environment,
in fact, when $\Delta$ goes to zero, the noise effect vanishes.
The autocorrelation function of the $c_j^{A(B)}$'s, given by 
$\langle\delta c(t)\delta c(0)\rangle=\frac{\Delta^2}{12}$, is time independent 
and this implies that its power spectrum is proportional to a $\delta$-function
centered on zero frequency. As a consequence, this kind of noise has a
characteristic time which is always much longer than the characteristic time of the
system-environment coupling.
 Therefore, the static disorder can be 
considered as representative of a non-Markovian noise.

\begin{figure}[htbp]
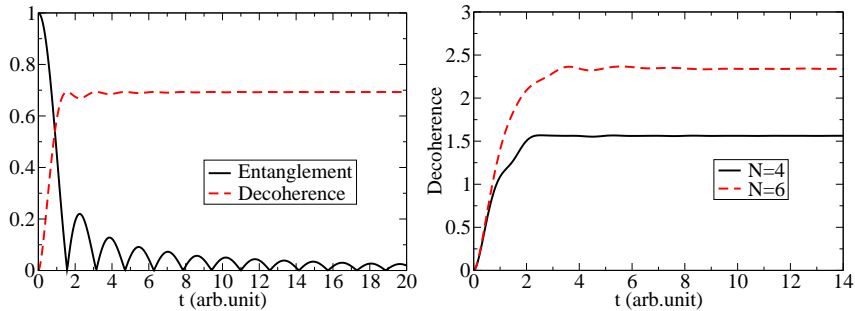

  \begin{centering}
   \includegraphics*[width=0.43\textwidth]{Negativity_vNentropy_N_2_Static_Common_Environ.eps}
   \includegraphics*[width=0.45\textwidth]{vNEntropy_N_4_6_Static_Common_Environ.eps}
  \caption{\label{fig1} Left panel: Entanglement (solid line) and decoherence (dashed line)
as a function of time for $N=2$. Right panel: decoherence time evolution for $N=4$ (solid line)
and $N=6$ (dashed line).}  \end{centering}
\end{figure}

In Fig.\ref{fig1} the time evolution of decoherence and entanglement are shown at $N=2,4,6$, 
for the input maximally entangled state $|\psi(t=0)\rangle=\frac{1}{\sqrt{N}}\sum_i^N |i_Ai_B\rangle$.
The entanglement,  here expressed in terms of negativity, as above explained
is calculated for the $N=2$ case only.
From the left panel of Fig.\ref{fig1} we see that the entanglement exhibits ESD and ER 
phenomena\cite{8},
corresponding to an oscillating behavior of the decoherence curve (here oscillations are much less
pronounced than for the entanglement case, because of the logarithmic dependence displayed in
Eq.(\ref{vN})). By increasing the number of sites (right panel), the saturation value of the 
decoherence increases because, as a consequence of the environment action, the system is described by
a statistical mixture involving a larger number of terms, and, for this kind of noise, 
we observe that these oscillations tend to disappear. 

\subsection{Dynamic noise}
In the dynamic disorder case the parameters $c_j^{A(B)}$'s are set equal to $\nu Q^{A(B)}_j(t)$ 
where $Q(t)$ is described by a stochastic process of the telegraph noise type, 
which flips randomly between the two values 1 and -1 
at a rate $\gamma$, and $\nu$ is the particle-environment coupling constant. 
The autocorrelation function of $Q(t)$, as well known, is given by
$\langle\delta Q(t)\delta Q(0)\rangle=e^{-2\gamma t}$ with the Lorentzian power spectrum
$\langle\delta Q\delta Q\rangle_\omega=4\gamma/\left(\omega^2+4\gamma^2\right)$.
The relative length of the two time scales $\tau_e\sim 1/\gamma$ (characteristic time of the
environment) and $\tau_c\sim 1/\nu$ (characteristic time of the particle-environment 
coupling) are critical for decoherence and entanglement. Indeed, 
$\tau_e<\tau_c$ is identified as the Markovian regime, while $\tau_e>\tau_c$ is identified as the 
non-Markovian regime\cite{8}.

\subsubsection{Markov case}

\begin{figure}[htbp]
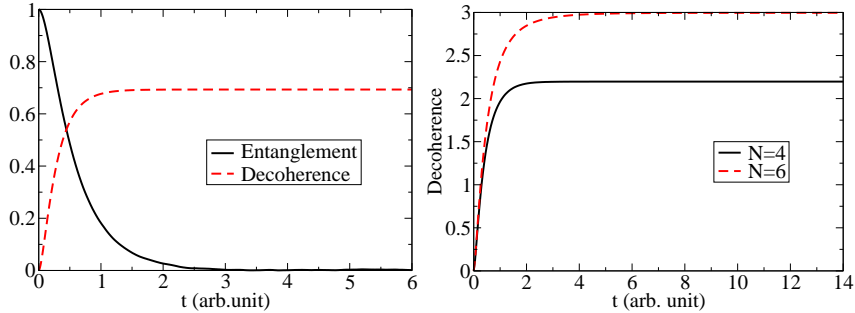

  \begin{centering}
   \includegraphics*[width=0.43\linewidth]{Negativity_vNentropy_N_2_RTN_markovian_Common_Environ_g5.eps}
   \includegraphics*[width=0.45\linewidth]{vNentropy_N_4_6_RTN_Markovian_Common_environment_g5.eps}
  \caption{\label{fig2} Left panel: Entanglement (solid line) and decoherence (dashed line)
as a function of time for $N=2$. Right panel: decoherence time evolution for $N=4$ (solid line)
and $N=6$ (dashed line). Here the Markov regime is obtained by setting $\tau_e/\tau_c=1/5$}  
\end{centering}
\end{figure}

In Fig.\ref{fig2} the results for the case $\tau_e/\tau_c=1/5$ are presented, for the same
initial state $|\psi(0)\rangle$ used in the previous section.
Unlike the case of static disorder, because of the short  time correlation of the environmental
noise, the degree of entanglement between the two-particles decays continuously and no
ESD and ER phenomena are observed. Decoherence is a monotonic function of time, and reaches 
asymptotic values larger and larger as the number of nodes increase.

\subsubsection{Non-Markov case}

\begin{figure}[htbp]
  \begin{centering}
   \includegraphics*[angle=270,width=0.49\linewidth]{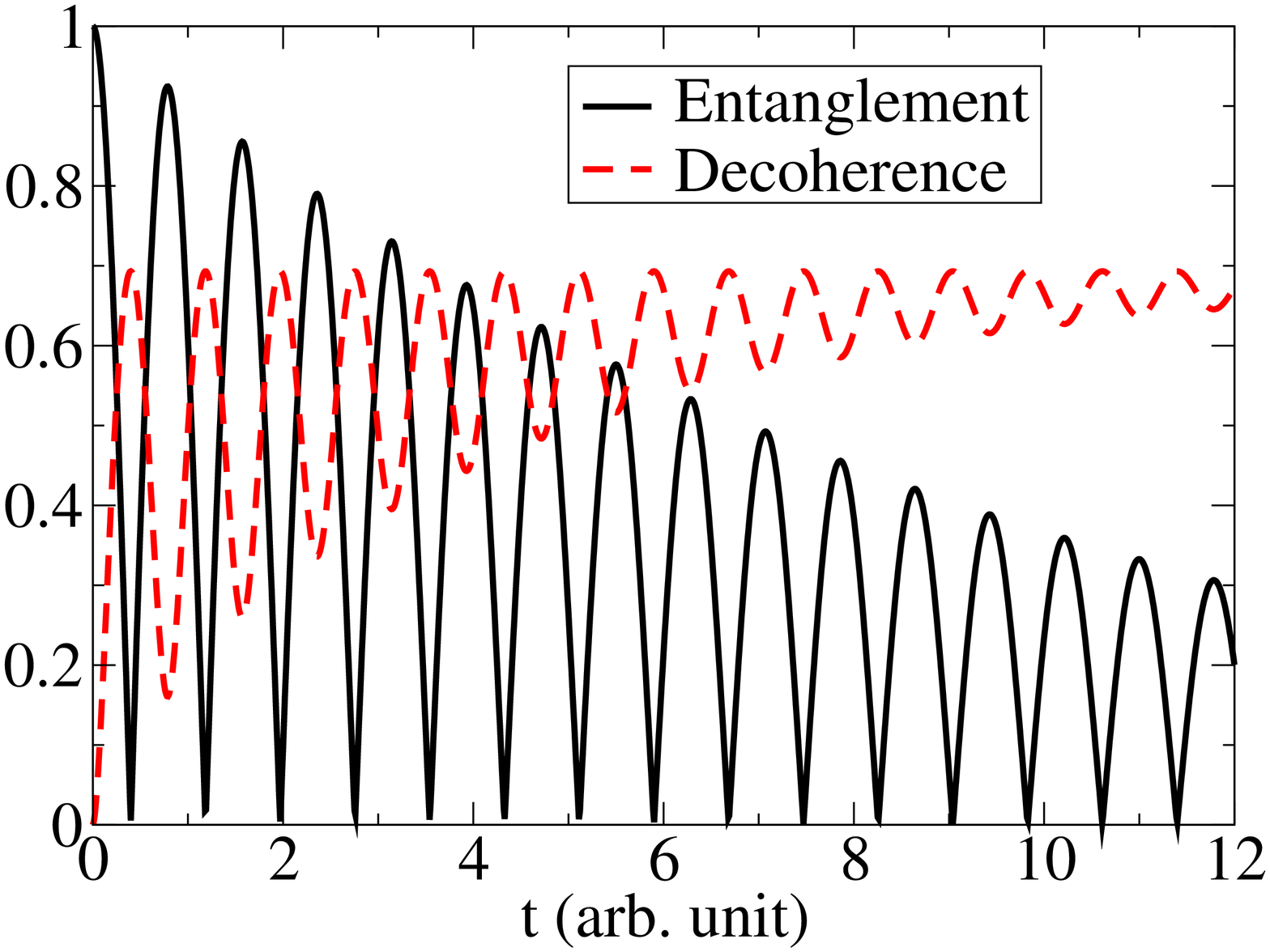}
   \includegraphics*[angle=270,width=0.49\linewidth]{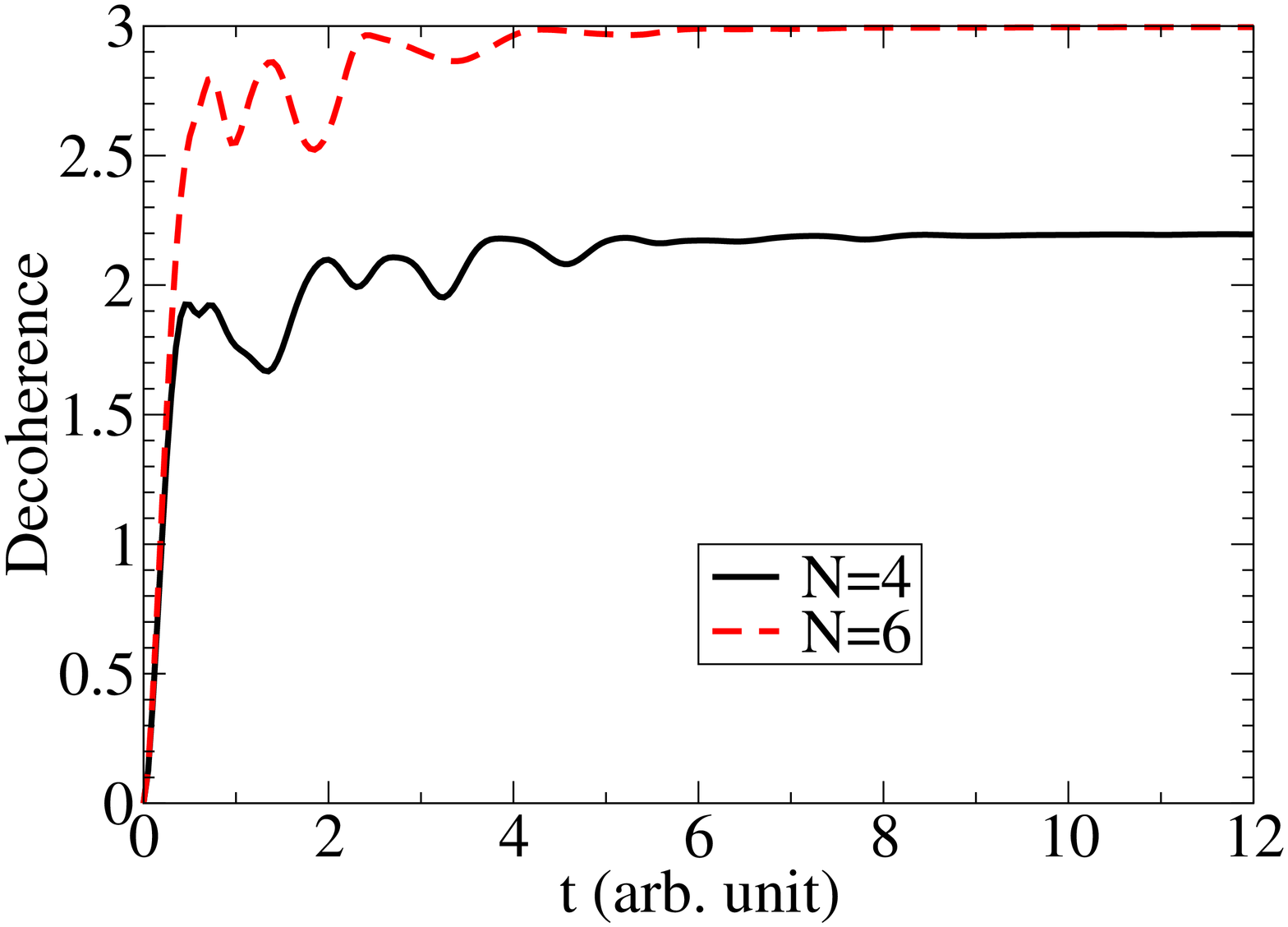}
  \caption{\label{fig3} Left panel: Entanglement (solid line) and decoherence (dashed line)
as a function of time for $N=2$. Right panel: decoherence time evolution for $N=4$ (solid line)
and $N=6$ (dashed line). Here the non-Markov regime is obtained by setting $\tau_e/\tau_c=10$.} 
\end{centering}
\end{figure}

The non-Markovian behavior has been simulated by setting the ratio $\tau_e/\tau_c=10$. The results 
are displayed in Fig.\ref{fig3}. For $N=2$ (left panel), the entanglement is a damped 
oscillating function of time, thus showing ESD and ER phenomena. Since the envelope 
is exponential we can assume that revival happens an infinite number of times, in agreement with
previous findings\cite{8}. Decoherence, unlike the Markovian case, presents evident oscillations
before reaching the saturation value. In particular we note that the local minima of 
decoherence correspond to the local maxima of entanglement, thus suggesting a strict
connection between the two quantities. In fact, in our model, we can attribute the
disentanglement of the two particles to the movement of the initial state $|\psi(0)\rangle$
towards mixed state. Such a relation can reasonably be thought to hold
also for graphs with $N>2$, where no rigorous evaluations of the entanglement can be performed.
Therefore, the oscillating behavior of decoherence observed for $N=4,6$ (right panel of 
Fig.\ref{fig3}) can be considered as an indicator of the presence of ER.

\section{Conclusions}

We have investigated the effect of a noisy environment on the time evolution of the entanglement
between two-quantum particles and of the loss of coherence of the two-particle system for 
CTQW in a one-dimensional ring lattice with periodic boundary conditions.
The two-particle initial state has always been chosen as a maximally entangled one.
The environmental noise has been simulated introducing  both static and dynamic disorder 
in the lattice. To this aim the transition amplitudes from one site to another have been 
assumed random and time independent (static case) or subject to a random telegraph
signal (dynamic case).

In the dynamic case, the persistence of the memory effect of the environment
is related to the switching rate of the random telegraph signal. Therefore two distinct regimes
can be identified according to the ratio between the correlation time of the environment and
the one characteristic of the system-environment interaction. If such a ratio is smaller than 1,
Markovian behavior can be assumed, on the contrary, values larger then 1 identify 
non-Markovian regime. The static disorder represents an environment whose memory effect 
can not be neglected at any finite time, therefore the characteristic time of its 
correlations is always larger than the one of the environment-system coupling. 
Thus we classify this kind of noise as non-Markovian. It should be noticed that such 
model is different from the one adopted in the literature\cite{17,18,19} where the non-Markovian noise 
is commonly represented only in terms of random telegraph signal.

For the Markovian case we found an exponential decay of the entanglement, while 
decoherence increases monotonically with time up to a saturation value, depending upon 
the number of lattice sites. No peculiar phenomena, like ESD or ER, are observed, 
in agreement with previous findings\cite{8}.

For the non-Markovian behavior we find that the dynamics of entanglement and decoherence
exhibits non-monotonic time dependence. In particular ESD and ER are detected\cite{7,8,9}
for $N=2$, and can even be supposed to be present for $N>2$ under some specific conditions
(dynamic noise). 
Indeed our approach suggests
that different kinds of noise leads to somehow different results. For static
disorder both entanglement and decoherence show less pronounced oscillations with
respect to the non-Markovian random telegraph noise for $N=2$, while for $N>2$ 
decoherence increase practically in monotonic way, thus suggesting that ESD and ER
phenomena are not present, unlike what is found for the dynamic noise case.

In conclusion, our analysis points out that the entanglement of a quantum system 
coupled to a classical source of noise is affected 
by the persistence of the memory effect of the
environment and by the intrinsic features of noise itself.
Therefore we believe that the extension of the present investigation to the ubiquitous case
of $1/f$ noise could be of great interest to get a better insight into the effect of
a classical noise on a quantum system\cite{20}.
Furthermore, in order to get a deeper understanding of how a classical environment
affects the appearance of quantum correlations, the study
of more complex  systems, including many-body effects and a higher number of degrees of freedom,
is required.
In this view, the extension of the approach here adopted to model many particles interacting 
with each other is not straightforward and will need the development of more sofisticated
and efficient numerical tools.

\section*{Acknowledgements}

The authors would like to thank M.G.A. Paris for fruitful discussions.

\end{document}